\documentclass[sigconf]{acmart}

\usepackage{algorithm}
\usepackage{algorithmic}
\usepackage{multirow}
\usepackage{caption}
\usepackage{array}

\usepackage{enumitem}

\setcopyright{acmlicensed}

\copyrightyear{2024}
\acmYear{2024}
\setcopyright{acmlicensed}\acmConference[ICAIF '24]{5th ACM International Conference on AI in Finance}{November 14--17, 2024}{Brooklyn, NY, USA}
\acmBooktitle{5th ACM International Conference on AI in Finance (ICAIF '24), November 14--17, 2024, Brooklyn, NY, USA}
\acmDOI{10.1145/3677052.3698637}
\acmISBN{979-8-4007-1081-0/24/11}

\begin{document}

\title{CustomizedFinGPT Search Agents Using Foundation Models}

\author{Felix Tian}
\authornote{Felix Tian finished this project as a RA at Columbia University during summer 2024.}
\affiliation{%
   \institution{Information Tech. \& Web Science, Rensselaer Polytechnic Institute}
   \city{Troy}
   \country{USA}
}

\author{Ajay Byadgi and Daniel Kim}
\affiliation{%
   \institution{Computer Science, Rensselaer Polytechnic Institute}
   \city{Troy}
   \country{USA}
}

\author{Daochen Zha}
 \affiliation{
   \institution{Independent Researcher}
   \city{San Francisco}
   \country{USA}
}

\author{Matt White}
 \affiliation{%
    \institution{Linux Foundation; Executive Director, PyTorch foundation; UC Berkeley}
   \city{Berkeley}
   \country{USA}
 }

\author{Kairong Xiao}
\affiliation{
  \institution{Roger F. Murray Associate Professor of Business, Business School, Columbia University}
  \city{New York}
  \country{USA}
}

\author{Xiao-Yang Liu Yanglet}
\authornote{Corresponding author. Email: XL2427@columbia.edu}
\affiliation{
  \institution{Electrical Engineering, Columbia University; Computer Science, Rensselaer Polytechnic Institute}
  \city{New York}
  \country{USA}
}



\begin{abstract}

Current large language models (LLMs) have proven useful for analyzing financial data, but most existing models, such as BloombergGPT and FinGPT, lack customization for specific user needs. In this paper, we address this gap by developing FinGPT Search Agents tailored for two types of users: individuals and institutions. For individuals, we leverage Retrieval-Augmented Generation (RAG) to integrate local documents and user-specified data sources. For institutions, we employ dynamic vector databases and fine-tune models on proprietary data. There are several key issues to address, including data privacy, the time-sensitive nature of financial information, and the need for fast responses. Experiments show that FinGPT agents outperform existing models in accuracy, relevance, and response time, making them practical for real-world applications.
  
\end{abstract}

\maketitle

\section{Introduction}

Large language models (LLMs) are transforming the financial sector by enabling efficient data analysis and decision-making support. These models, commonly referred to as financial LLMs (FinLLMs), process large datasets and offer general insights. However, existing FinLLMs like BloombergGPT \cite{wu2023bloomberggpt} and FinGPT \cite{yang2023fingpt,liu2023fingpt} can not provide customized advice to individual users and institutions, particularly when dealing with proprietary or personal data.

There is a growing demand for customizable solutions that can address the unique requirements of different users. For individuals, personalized financial guidance is essential in areas like retirement planning or investment strategy. For institutions, the need includes advanced analysis of proprietary datasets, and trading records and portfolio management, while ensuring data security.

Therefore, in this paper,\textbf{\emph{ we propose developing two versions of FinGPT agents: one customized for individuals and another customized for institutions.}} Each version is optimized to meet user-specific needs, integrating data from diverse sources while ensuring privacy and real-time updates. By using Retrieval-Augmented Generation (RAG) and model fine-tuning, the FinGPT agents provide customized financial insights, while ensuring data privacy, information freshness, and response speed.

\vspace{-5pt}

\begin{figure*}[t]
    \includegraphics[width = 0.8\textwidth] {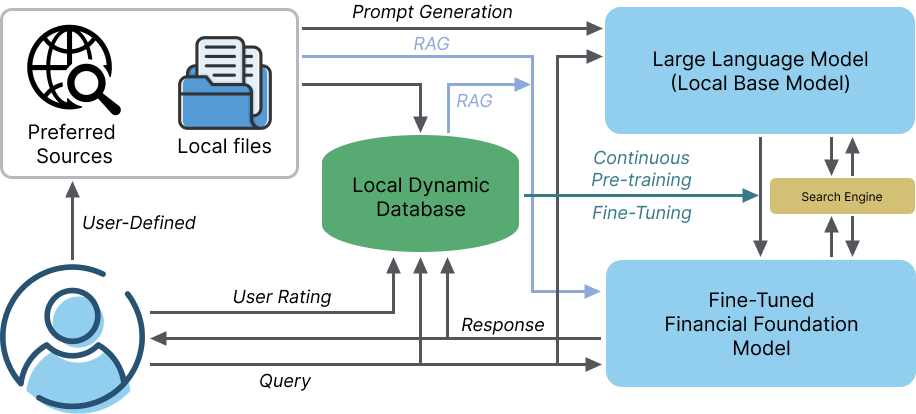}
    \centering
  \caption{The overall dataflow of FinGPT agents. A user submits a query, while an agent retrieves relevant data from preferred sources, search engines, local files, or a dynamic vector database. The LLM or fine-tuned financial foundation model (FFM) processes the prompts, refines them, and utilizes Retrieval-Augmented Generation (RAG) to generate responses. Both users' feedback and generated responses are then stored into the local dynamic vector database for future interactions.}
  \Description{Overall Information Flow}
  \label{fig:teaser}
  \vspace{-0.1in}
\end{figure*}

\vspace{5pt}

Developing such FinGPT agents is non-trivial due to several challenges. \textbf{\emph{(i) Privacy.}} For both individual users and institutions, privacy and data security are paramount, as the agent needs access to sensitive financial information. Ensuring the security of this data while providing accurate responses is a significant challenge. \textbf{\emph{(ii) Time-sensitivity.}} Financial data is time-sensitive. Ensuring the timely processing and integration of the most up-to-date information efficiently poses technical challenges. \textbf{\emph{(iii) Agent Response Time.}} Users will not wait for more than a dozen seconds for a response. Ensuring high-quality responses while maintaining timeliness is a challenging task. Through addressing the above challenges, we make the following contributions:
\begin{itemize}[leftmargin=*]
    \item We explored customizable FinGPT agents for individuals and institutions. In contrast to generic financial advices provided by current FinLLMs, customizable FinGPT agents provide customizable solutions based on unique requirements of different users, such as proprietary or personal data.
    \item We proposed two highly customized FinGPT agents: one for individuals and one for institutions. Our design ensures that both agents preserve data privacy while providing real-time processing of financial data and responses. We also designed and developed tailored Graphical User Interfaces (GUIs) for individual users and institutional users, respectively.
    \item Experiments on financial questions demonstrate that the proposed FinGPT search agents outperform existing LLMs in terms of accuracy, data freshness, and response time, showing their potential in real-world applications.
\end{itemize}

\section{Related Work}

\subsection{Web Agents}

Web agents refer to the tools that perform intricate tasks across a variety of real-world websites, often leveraging language models to process web content. A recent example of this is MIND2WEB~\cite{chen2023mind2web}, a generalist web agent designed to manage the dynamic nature of websites. This process is twofold: firstly, a compact language model filters webpage elements, secondly, a larger language model predicts actions based on those elements. Another notable work in this field is WebVoyager~\cite{he2024webvoyager}, a multi-modal web agent. WebVoyager navigates real-world websites by utilizing both visual and textual elements. It significantly enhances its decision-making process by utilizing screenshots and web elements. This improvement enhances performance on complex tasks, such as interacting with dynamically evolving websites. Our approach differs from the existing web agents since we focus specifically on the financial sector and enable users to input proprietary or personal data. This design allows our agents to offer highly customized solutions.

\subsection{Information Retrieval (IR)}

Information retrieval (IR) focuses on extracting relevant information from data. Recently, LLMs have been explored in IR to enhance query formulation and relevance ranking~\cite{zhu2023large}. Approaches such as InPars~\cite{bonifacio2022inpars} propose utilizing LLMs as synthetic data generators, addressing the limitations of general-purpose IR datasets. This allows retrieval models to be fine-tuned with minimal supervision while still achieving strong performance. SimplyRetrieve~\cite{ng2023simplyretrieve} distinctly separates the roles of LLMs and retrievers, with LLMs interpreting context and retrievers storing knowledge. This strategy optimizes the IR process by reducing reliance on LLMs and enhancing output interpretability, while also mitigating hallucination issues common in other generative models. Our FinGPT agent can be viewed as an IR tool specifically designed for financial data.

%Information retrieval concerns with fetching information online or offline. This process is the core of any customizable agents. A survey done by Zhu et al.~\cite{zhu2023large} provided an insight of how LLMs can personalize IR by improving query formulation and ranking relevance, which is directly relevant to the FinGPT Search Agent. This demonstrates the pivotal role of advanced IR systems in personalizing web agents.

%Recent approaches like InPars~\cite{bonifacio2022inpars} propose using LLMs as synthetic data generators to overcome limitations of general-purpose IR datasets. Their method generates synthetic question-document pairs in few-shot, allowing retrieval models to be fine-tuned with minimal supervision and still obtain better-than-base-model performance. SimplyRetrieve~\cite{ng2023simplyretrieve} introduces Retrieval-Centric Generation (RCG), which explicitly separate the roles of LLMs and retrievers. LLM  focuses on interpreting context, while the retriever memorizes knowledge. This approach optimizes the IR process by reducing reliance on large LLMs and improving interpretability of outputs. It mitigates hallucination compared to other generative models.

\subsection{Large Language Models (LLMs)}

LLMs, built on the Transformer \cite{vaswani2017attention} architecture excel at understanding and generating human language. The attention mechanism within these models enables them to capture contextual dependencies across input sequences, making them suitable for a variety of language tasks.

Early LLMs, such as GPT-1~\cite{radford2018_improving}, demonstrated their potential by generating coherent text. Subsequent models, like GPT-3~\cite{brown2020_language_model} and LLaMA~\cite{touvron2023_LLaMA}, further expanded model capacity, enabling broader applications. In finance, BloombergGPT~\cite{wu2023bloomberggpt} marked a significant step forward by integrating domain-specfici data with general-purpose information. However, despite these advancements, existing works are still unsatisfactory for providing customized financial insights and much more focused on generalized interactions.

\subsection{LLM Finetuning Methods}
Fine-tuning adapts generic LLMs to specific tasks or domains. Recent techniques like Low-Rank Adaptation (LoRA)~\cite{hu2022_lora} and (QLoRA)~\cite{dettmers2023qlora} improves efficiency and reduces memory usage of tuning. As shown in these works \cite{hu2022_lora} \cite{dettmers2023qlora} \cite{liu2024fingpt}, these methods indeed specializes LLMs in financial applications. However, fine-tuning alone is insufficient for addressing the time-sensitiveness~\cite{liu2024dynamic} of financial data as this data constant updates. To overcome this, we combine fine-tuning with Retrieval Augmented Generation (RAG).

\subsection{Retrieval-Augmented Generation (RAG)}

RAG integrates retrieval mechanisms into generative models, ranging from simple keyword matching to advanced neural retrieval models~\cite{zhao2024survey}. It represents a cost-efficient way to incorporate the latest information into LLMs since it does not require updating model weights. Liu et al. \cite{liu2023dynamic} explored the integration of RAG with fine-tuned models, demonstrating improved performance in generating contextually rich and precise outputs for domain-specific applications. In Zhang et al.'s work~\cite{zhang2023retrieval} on financial sentiment analysis, they combined RAG with instruction tuning. They identify two challenges: the mismatch between LLM pre-training objectives and sentiment analysis, and the lack of sufficient context in short financial texts like tweets. To overcome this, they use a retrieval module that gathers relevant external financial data, enriching the context for sentiment predictions. Their approach improves performance by 15-48\% in accuracy and F1 score over traditional models.

Although RAG combined with fine-tuning has been explored in previous work, it has not been used for customizing LLMs. Our work leverages RAG to provide a customized experience for individual users and institutions, supported by a GUI.

\subsection{LLM Customization Methods}
Although existing studies have adapted LLMs to vertical domains, they are not capable of providing customized responses in a user-centric manner~\cite{esteva2021_medical_ai,chalkidis2020legal,wu2023bloomberggpt}. In particular, existing FinLLMs, such as BloombergGPT~\cite{wu2023bloomberggpt}, only offer generalized insights rather than tailored financial advice. Our work on the FinGPT agents aims to push the boundaries of LLM customization in the financial sector. By integrating proprietary financial data, fine-tuning models with institution-specific datasets, and employing RAG, we can provide highly personalized and contextually relevant financial insights.

\section{Local Customizable FinGPT Agent for Individual Users}

\subsection{Overview}

\begin{figure*}[t]
   \centering
   \includegraphics[width=0.65\textwidth]{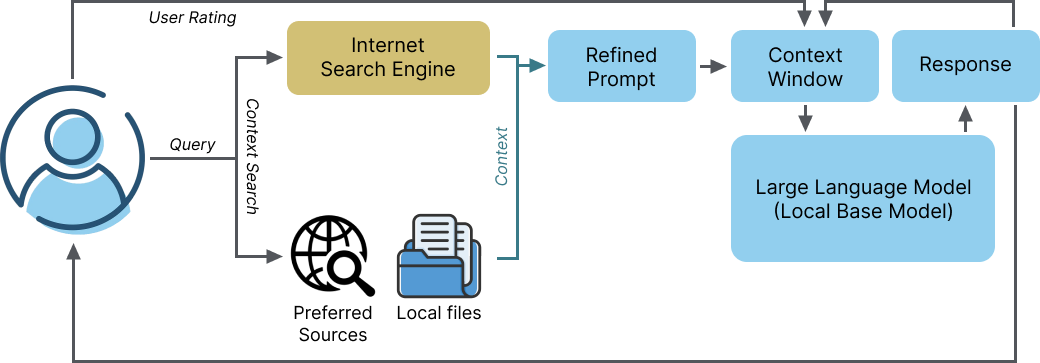}
    \caption{An overview of customized FinGPT agent for individual users.}
    \label{fig:2}
    \vspace{-0.15in}
\end{figure*}

We present an overview of the customized version of the FinGPT search agent for individual users, as shown in Fig. \ref{fig:2}. The agent runs locally on users' computers. Users specify preferred web sources (e.g., Web URLs and APIs) and allow access to local files. In particular, the agent utilizes search engines like Google to access web data, including up-to-date financial information.

\textbf{User End.}
From the user's end, the FinGPT agent serves as a search agent and a financial assistant:
\begin{itemize}[leftmargin=*]
    \item A user submits a query. 
    \item The FinGPT agent generates a response to the user.
    \item The user provides feedback that improves future interactions.
\end{itemize} 

\textbf{Model End.}
From the model's end, the FinGPT agent processes the user’s inquiries as follows:
\begin{itemize}[leftmargin=*]
\item The agent receives users' queries. 
\item The agent retrieves relevant information and context from three main sources:
\begin{itemize} 
\item Preferred sources specified by a user as authentic information, which are accessed via Web APIs. 
\item Local files such as PDFs and Excel sheets stored on the user's computer are also retrieved.
\item Relevant web data gathered by a search engine. 
 \end{itemize} 
\item The agent refines the user’s query based on the gathered data and generates a contextual-rich prompt.
\item The refined prompt and retrieved data are put into a context window. The backbone LLM model processes this contextual information and generates a response.
\item The response is sent back to the user and routed back through the context window, ensuring ongoing improvement in the agent’s accuracy and relevance. 
\end{itemize} 
By integrating data from multiple sources, the FinGPT agent effectively acts as a robo-advisor. This allows users to receive personalized financial insights and recommendations, leveraging a broad spectrum of data inputs for accurate and relevant advice. 

\subsection{User Preferences and Search Engine}

Incorporating user preferences includes incorporating a user-specified web link list and the user's local files. Open domain search are utilized to ensure that the generated response is up-to-date using keywords extracted from the refined prompt.

\textbf{Setting User-Preferred Web Lists}: Users specify preferred web sources (e.g., Web
URLs and APIs) through the search agent’s UI. The agent prioritizes retrieving information from these links before retrieving information via search engines. The Reader library by jina-ai is used to convert URLs into a more LLM-friendly format \cite{jina_ai_reader}. The reader appends ``https://s.jina.ai/'' to the front of a search query, allowing it to automatically fetch the top 5 search results. It then visits each URL and appends ``https://r.jina.ai/'' before each link. The second appended URL parses each website. Different from traditional web source fetching, Reader fetches the content of target websites instead of only returning the metadata of the fetched URLs provided by the search engine API.

\textbf{Fetching and Parsing Local Files}: Users specify a file path through the agent's UI, and the agent utilizes the Marker library to parse the files stored in that path. The Marker library \cite{pdf_marker} converts PDFs to markdown format for easier processing.

\textbf{Preferred Web Sources}: By focusing on incorporating user preferences, fetching local files, and initiating web searches, the FinGPT Search Agent gathers relevant and up-to-date information tailored to the user's needs, which enhances the relevance and accuracy of the financial advice provided. For example, our FinGPT Search Agent can leverage a search engine to find the context of an user query and perform more accurate sentiment analysis~\cite{zhang2023retrieval}.

\subsection{Refined Prompt and Long Context Window}

Once the search agent has retrieved relevant context, the agent utilizes this information to refine the user's query into a precise prompt and prepares the context for the model. 

Below is an example refined prompt for a user query.

\textbf{User Inquiry}: \textsf{Predict the next quarter's income for our company in the AI sector.}

\textbf{Refined Prompt}: \textsf{Based on the latest financial reports, market trends, and recent news articles about the AI sector, predict the next quarter's income for our company.}

The long context window includes all relevant data that the LLM will use to generate the response, including the refined prompt. This step makes the agent ``customizable'' to the user.

\subsection{Generating Responses \& Updating Context}

The models responses with all available information in the context window, and then updates the context windows to ensure continuity in subsequent interactions. All data processing and model operations occur locally for privacy.
Users may evaluate the responses and provide feedback to the agent. The feedback is input into the context window. By continuously updating the context window, the agent maintains a history of interactions. This continuity ensures that each subsequent response is informed by previous interactions.

\section{Local Customizable FinGPT Agent for Institutions}

\subsection{Overview}

\begin{figure*}
    \includegraphics[width = 0.8\textwidth]
    {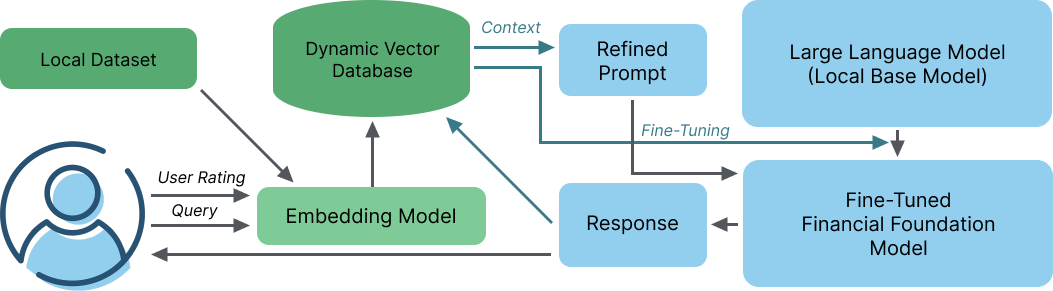}
    \centering
  \caption{An overview of customized FinGPT agent for institutional users.}
  \Description{Enjoying the baseball game from the third-base
  seats. Ichiro Suzuki preparing to bat.}
  \label{fig:3}
\end{figure*}

We present an overview of the customized version of the FinGPT search agent for institutional users, as shown in Fig. \ref{fig:3}.
 will be integrated into an institution’s existing air-gapped infrastructure. An institution's proprietary data are embedded via any choice of local embedding methods into a built-in dynamic vector database, which is used to continuously pre-train and fine-tune a base model.

\textbf{User End.}
From the institution's end, the FinGPT agent serves as a secure and efficient tool for financial analysis:
\begin{itemize}[leftmargin=*]
\item A user submits a query or request for financial analysis. 
\item The FinGPT agent generates a response based on the refined prompt and context. 
\item The user receives the generated response.
\item The user may provide feedback to the search agent.
\end{itemize}

\textbf{Model End.}
An institution may store its proprietary datasets (in JSON format) in the dynamic database. We use such proprietary data to continuously pre-train and fine-tune an LLMrom the model's end, the FinGPT agent processes users' inquiries as follows:
\begin{itemize}[leftmargin=*]
\item The user submits a financial query.
\item The agent retrieves relevant embedding vectors from the dynamic database.
\item The user’s prompt is refined using the embedding vectors.
\item The refined prompt and retrieved data are put into a context window. 
\item The model uses contextual information to generate a response.
\item The generated response is sent to the user and fed back into the dynamic vector database. The user rating is sent to the database as well, updating the context and improving the quality of future interactions. 
\end{itemize}
Integrating data from multiple sources  their data to enhance the model’s responses through both fine-tuning and RAG, without compromising data security. 

\subsection{Embedding Model}

The embedding model transforms raw data from the local dataset and user queries into embedding vectors. 

%These embeddings are essential for creating a contextually rich database. 

\textbf{Embedding Generation}: The search agent uses Mongo NoSQL Database as its dynamic database. Institutions set up the connection between the agent and existing data sources via the agent's GUI. The agent fetches the dataset and stores it in the dynamic database. The embedding model then converts the dataset into dense vectors. Various models can be used for generating embeddings. We use BERT \cite{devlin2019bert} as an example to describe the process:
\begin{itemize}[leftmargin=*]
    \item The BERT model is loaded from a pre-trained checkpoint.
    \item The text data is tokenized into individual tokens. 
    \item The tokens are converted into input IDs as numerical representations. These input IDs are fed into the BERT model.
    \item The BERT model processes the input IDs and generates dense vector embeddings, which represent the input text in a high-dimensional space, capturing its semantic meaning.
\end{itemize}

\textbf{Updating the Dynamic Database}: Generated embeddings are stored back into MongoDB, creating a dynamic vector database that the FinGPT search agent can efficiently query.

\subsection{Dynamic Database}

The dynamic database stores raw data and embeddings, and updates in real-time from the user's interactions with the agent. It ensures efficient storage, retrieval, and updating of vector embeddings generated from institutions' presumably large datasets and user queries.

Mongo NoSQL Database is used due to its ability to handle unstructured data and support of vector searches. MongoDB's flexibility and scalability make it suitable for integrating with LLMs. As shown in~\cite{hoesada2023enhancing}, the process for storing, embedding, and retrieving the data can be easily streamlined in MongoDB's NoSQL databases.

Each embedding is inserted into a specific collection within the database. When a query is made, query embeddings are generated using the same pre-trained embedding model. These query embeddings are used to perform a vector search within the MongoDB collection to retrieve relevant embeddings using the vector search ability provided by MongoDB. 
This approach allows for efficient and accurate retrieval of information for the model.

\textbf{Ensuring Security and Privacy}: Given that the FinGPT agent operates within an air-gapped infrastructure, the dynamic vector database must also adhere to strict security and privacy standards, operating entirely locally.

\subsection{Fine-tuning LLM Base Model}

The base model used by the agent can be fine-tuned to create a local, institution-specific LLM.
The data stored in the dynamic vector database is organized into training batches pre-tuning. During each training epoch, the model processes each batch of data, generating predictions based on the input data. The difference between the predicted outputs and the true values (targets) is calculated using a loss function. Next, the gradients of the loss with respect to the model parameters are computed. These gradients indicate how the model parameters need to be adjusted to minimize the loss. Model parameters are then updated using these gradients via gradient descent, according to a specified learning rate. This process is repeated for all batches in the dynamic vector database and across multiple training epochs.

\subsection{Response Generation \& Context Enrichment}

The user's query is refined into a more contextually rich and relevant prompt by the model. The model then reads this refined prompt, and utilizes RAG with data from the dynamic vector database to further enhance the generated response from the fine-tuned model. 

\textbf{Ensuring Continuous Improvement}.
Generated responses, refined prompts, and user feedback are embedded and fed back into the dynamic vector database. It ensures a subsequent response is informed by previous interactions and improves responses' accuracy. This continuous loop of user-agent interaction ensures that the FinGPT search agent remains highly relevant and accurate, providing institutions with the best possible financial insights and/or analysis.

\begin{figure*}[t]
    \centering
    \includegraphics[width=1.02\textwidth]{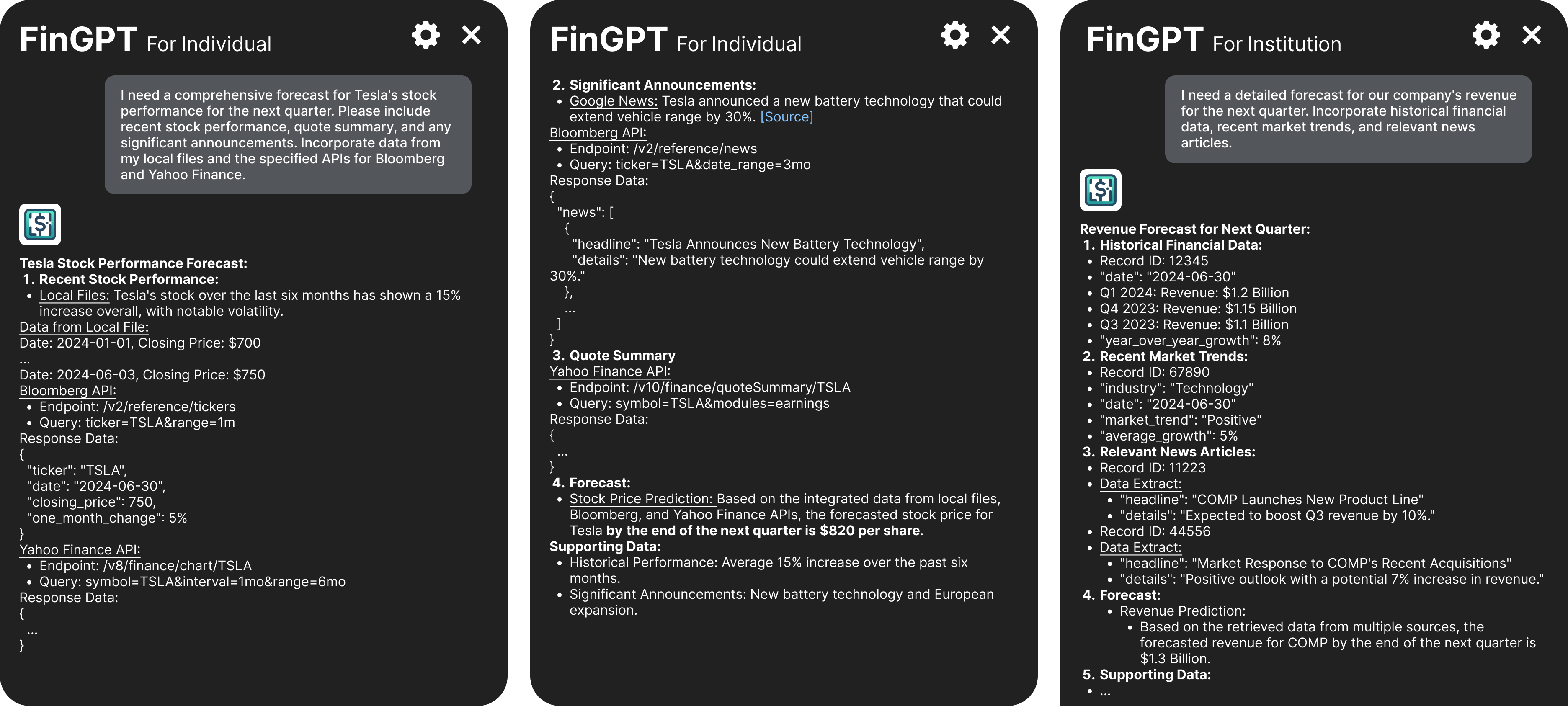}
    \hspace{0.3in}     \caption{Screenshots of the customized FinGPT search agents for individual users (left \& middle) and institution users (right).}
     \label{fig:4}
     \vspace{-0.15in}
\end{figure*}

\section{Graphic User Interface (GUI)}

To facilitate ease of use of the proposed FinGPT search agents, we developed tailored GUIs for both individual and institutional users.

\subsection{GUI for Individual Users}

The two screenshots on the left-hand side of Fig.~\ref{fig:4} show a sample interaction from individual version. All buttons are labeled and displayed directly in the main window, making the UI straightforward to use even for those who aren't familiar with computers. The UI consists of a main window displaying the current interactions with the model, a source button, a clear button, a button for setting user preferences, and a white mode toggle. The source button displays all sources used for generating current responses and clears when starting a new conversation. The user preferences button opens up a pop-up for adding web links, API endpoints, and paths to local files. The agent's built-in default settings allows it to be used without any advanced configuration. However, if users want to, they may do it inside Settings.

As demonstrated by the two screenshots, the search agent customized for individuals, upon receiving a query, retrieves data from user-specified API endpoints and local files, lists them out, and draws a conclusion with supporting data.

\subsection{GUI for Institutional Users}
The screenshot for institutional users is shown on the right side of Fig.~\ref{fig:4}. It remains largely the same as the version customized for individuals, with an additional pop-up from the main window for inputting local datasets. Institutions need to explicitly write code in the provided field to transfer data. This gives institutions full control and customization over the data flow. The search agent's base model may be fine-tuned once the dynamic database is saturated with relevant data via another pop-up. The agent notifies its user from the fine-tune pop-up once the fine-tuning is complete.

Upon receiving a query, the search agent customized for institutions automatically retrieves relevant financial data and news articles from the dynamic vector database, lists them out, and presents a forecast with supporting data.

\section{Performance Evaluation}

In this section, we evaluate the proposed FinGPT search agents by addressing two research questions:

\begin{itemize}[leftmargin=*]

\item \textbf{RQ1}: How effective are the FinGPT search agents in terms of accuracy and data freshness?
\item \textbf{RQ2}: How efficient are the FinGPT search agents in terms of inference time and response time?

\end{itemize}

\subsection{Experimental Settings}

\subsubsection{Datasets} For the \emph{individual version}, we test s:
\begin{itemize}[leftmargin=*]
\item \textbf{Financial Questions:} This dataset includes 108 financial quantitative questions are divided into 54 \emph{easy} questions and 54 \emph{hard} questions.
Easy questions are straightforward numbers or facts, such as stock prices or percentage changes, while hard questions require analysis and explanation, such as asking how markets will react to an event. 
\item \textbf{Web Questions:} This dataset involves 200 web-page-specific questions, e.g., asking to explain an article headline, and inquiring about featured stock prices, where 111 of them are based on \emph{Yahoo Finance}, and 89 are based on \emph{Bloomberg}. They are also divided into 88 \emph{easy} and 112 \emph{hard} questions. 
\end{itemize}

%The details of the above datasets are provided in this link\footnote{\url{https://github.com/AjayByadgi/FinGPT-Individual-Agent-Performance}}. 

For the \emph{institution version}, we focus on the following five datasets:
\begin{itemize}[leftmargin=*]
\item \textbf{EMIR} \cite{finos_emir_rag}: The European Market Infrastructure Regulation (EMIR) dataset is a part of the broader effort to facilitate RAG that can handle complex regulatory data. It includes \emph{abbreviations} and \emph{definitions}, where the former aims to identify the full names of abbreviations and the latter aims to correctly explain regulations, standards, and concepts of terms.
\item \textbf{ESMA} \cite{finos_emir_rag}: It is scraped from the European Securities and Markets Authority (ESMA), and its questions are about \emph{abbreviations}.
\item \textbf{NER} \cite{individual_performance_2024}: This dataset tests a model's ability to recognize Named Entity Recognition (NER), which aims to interpret and recognize EMIR regulations given \emph{fragments} of the regulations.
\item \textbf{Link Retrieval} \cite{individual_performance_2024}: This dataset is comprised of 100 documents retrieved from EMIR's website. The aim is to test the models and the search agent's ability to find their \emph{corresponding URLs}. 
\item \textbf{FinanceBench} \cite{islam2023financebench}: This dataset comprises questions about reports filed by publicly traded companies with corresponding answers and evidence strings. The questions can be categorized into three types. \emph{Domain-relevant} questions address industry-specific or domain-specific requirements. \emph{Novel-generated} questions focus on concepts and ideas (e.g. emerging trends, interdisciplinary approaches that test reasoning). \emph{Metrics-generated} questions focus on the analysis of quantitative data and performance metrics.
\end{itemize}

\subsubsection{Baselines} 

We compare the customized FinGPT search agents with state-of-the-art LLMs, including \textbf{GPT-3.5} (often known as ChatGPT), \textbf{GPT-4o} (OpenAI's new flagship model designed for real-time multimodal interactions with faster response times and enhanced performance)~\cite{openai2024gpt4o}, and \textbf{LLaMA 3}, the largest model in~\cite{touvron2023_LLaMA}, trained on 65 billion tokens from public datasets.

\subsubsection{Metrics} 
We evaluate performance using accuracy, where each response receives a score. A response is given one of the following scores based on its correctness and data freshness, and the accuracy is calculated by averaging these scores.
\begin{itemize}[leftmargin=*]
\item\textbf{1} is given if the response is relevant, up-to-date, and correct. For example, for the question ``Who is Apple's CEO?'', a score of 1 will be given if the agent answers ``Tim Cook''.
\item \textbf{0.5} is given if the response is relevant and correct but outdated. For instance, the agent will be given a score of 0.5 if it answers ``Steve Jobs'' for the question above.
\item \textbf{0} is given if the response is either irrelevant or incorrect. For the question "Who is Apple’s CEO?", the score will be 0 if the agent answers ``Microsoft is headquartered in Redmond, Washington'' (irrelevant) or ``Satya Nadella'' (incorrect).
\end{itemize}

In addition, to understand the cost of the FinGPT search agents, we report \textbf{response time}, which measures the total time from receiving the query to generating the response.

%ORGANIZING:
%For individual testing, for Table 1 and 2, we are using the dataset designed by Ajay. Explanation and justification will be provided by him. 
%For institution testing, we look at table 3 figure 5 and 6. For table 3, we are using EMIR and EMSA:
%https://github.com/finos-labs/emir-specific-rag/blob/main/benchmarking/Report%20of%20benchmarking%20EMIR%20regulation%20data.pdf

%https://github.com/finos-labs/emir-specific-rag#license 
%https://www.esma.europa.eu/press-news/esma-news/esas-published-second-batch-policy-products-under-dora

%This is for Financial Regulations. For figure 5 and 6, we are using SEC 10K filings:
%https://arxiv.org/abs/2311.11944
%https://github.com/patronus-ai/financebench

\begin{table*}[t]
\centering
\caption{Performance on web tasks.}
\label{tab:model_comparison}
\begin{tabular}{lcc|cc|cc}
\toprule
\multirow{2}{*}{\textbf{Model/Agent}} & \multicolumn{2}{c}{\textbf{Financial Indices}}  & \multicolumn{2}{c}{\textbf{Web Questions (Yahoo Finance)}} & \multicolumn{2}{c}{\textbf{Web Questions (Bloomberg)}} \\ 
\cmidrule{2-7}
 & Easy & Hard & Easy & Hard & Easy & Hard \\ 
\midrule
\textbf{GPT-3.5} \cite{openai2023gpt3.5} & $0.22$ & $0.21$ & $0.13$ & $0.12$ & $0.14$ & $0.11$ \\ 
\textbf{FinGPT Search Agent} (GPT-3.5 as base) & $\textbf{0.90}$ & $\textbf{0.86}$ & $\textbf{0.92}$ & $\textbf{0.89}$ & $\textbf{0.94}$ & $\textbf{0.91}$ \\ 
\midrule
\textbf{GPT-4o} \cite{openai2024gpt4o} & $0.26$ & $0.34$ & $0.22$ & $0.23$ & $0.23$ & $0.25$ \\ 
\textbf{FinGPT Search Agent} (GPT-4o as base) & $\textbf{0.92}$ & $\textbf{0.93}$ & $\textbf{0.95}$ & $\textbf{0.94}$ & $\textbf{0.91}$ & $\textbf{0.93}$ \\ 
mrule
\end{tabular}
\end{table*}

\begin{table*}[t]
\centering
\caption{Performance on regulatory tasks.}
\label{tab:model_comparison_2}
\begin{tabular}{llcccccc}
\toprule
\multicolumn{2}{l}{\textbf{Regulatory Tasks}} & \textbf{GPT4} \cite{openai2023gpt4} & \textbf{GPT-4o} \cite{openai2024gpt4o} & \textbf{GPT-3.5} \cite{openai2023gpt3.5} & \textbf{Mistral} & \textbf{Llama3} \cite{llama3modelcard} & \textbf{FinGPT Search Agent} \\ 
\midrule
\multirow{2}{*}{\textbf{EMIR} \cite{finos_emir_rag}} & Abbreviations & 0.70 & 0.78 & 0.53 & 0.66 & 0.78 & \textbf{0.85}\\ 
 & Definitions & 0.90 & 0.96 & 0.32 & 0.66 & 0.54 & \textbf{0.98} \\  
\midrule
\textbf{ESMA} \cite{finos_emir_rag} & Abbreviations & 0.70 & 0.78 & 0.54 & 0.66 & 0.75 & \textbf{0.80}\\ 
 & Definitions & 0.9 & 0.96 & 0.32 & 0.66 & 0.54 & \textbf{0.98}\\ 
\midrule
\multirow{1}{*}{\textbf{Link Retrieval}} & Laws & 0.25 & 0.37& 0.21 &0.23 & 0.05 & \textbf{0.70}\\
\midrule
\multirow{6}{*}{\textbf{NER}} 
 & Organizations & 0.86 & 0.61 & 0.69  & 0.61 & 0.65 & \textbf{0.94}\\ 
 & Legislations & 0.96 &0.94  &  0.92&  0.96& 0.98 & \textbf{0.98}\\
 & Dates & 0.98  &  0.90 &  0.82 & 0.98 &  0.94 & \textbf{0.98} \\
 & Monetary values &  0.98&  0.96&  0.86&  0.98 & 0.96 & \textbf{0.98} \\
 & Statistics &  0.96 &  0.84& 0.67 &  0.94 & 0.86& \textbf{0.98} \\
\bottomrule
\end{tabular}
\vspace{-0.15in}
\end{table*}

\subsection{Effectiveness of FinGPT Search Agent (RQ1)}

\subsubsection{Results of individual version}

We evaluate the individual version of the FinGPT search agent against the corresponding base model in Table~\ref{tab:model_comparison} and make the following observations.

The individual version of the FinGPT search agent significantly outperforms the base models across all datasets, achieving an accuracy of 0.86 to 0.95. The base model only achieved an accuracy of 0.11 to 0.26. These results demonstrate the effectiveness of using auxiliary sources as context and employing RAG. Improvements are more pronounced on web questions. GPT-3.5 has an accuracy of around 0.20 on web questions and financial indices, while the search agent maintains around 0.90 accuracy on both financial indices and web questions. Base models may struggle to answer web questions correctly without retrieving web pages, while FinGPT can do so using RAG.

Using a stronger base model improves the FinGPT search agent's performance, as shown by the FinGPT Search Agent with the GPT-4o base consistently outperforming the one with the GPT-3.5 base. This suggests that the FinGPT search agent can be easily upgraded by integrating new LLMs, demonstrating its flexibility.

\begin{figure}[t]
    \centering
     \includegraphics[width=0.48\textwidth]{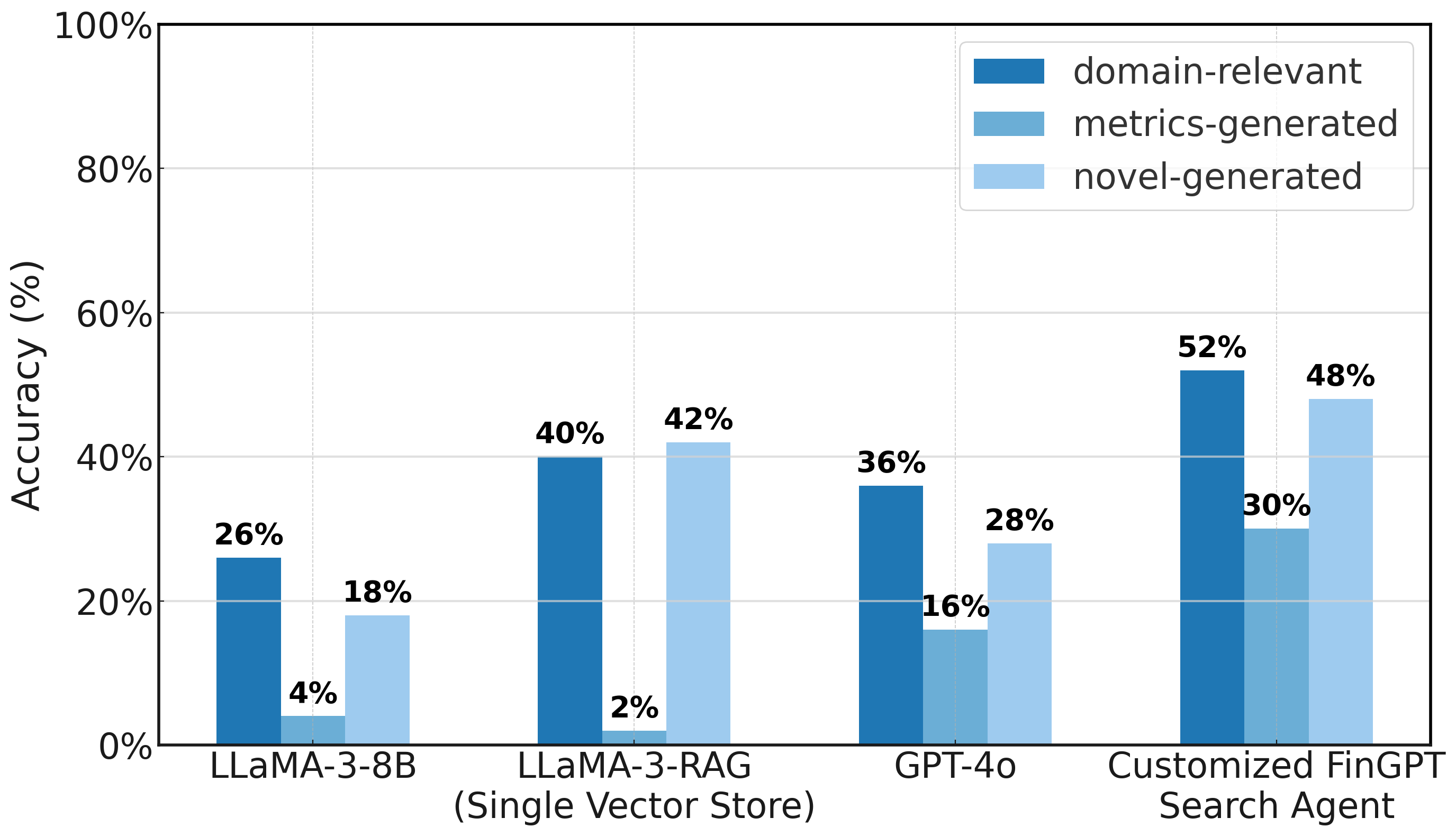}
    \caption{Performance on the FinanceBench dataset \cite{islam2023financebench}.}
    \label{fig:5}
    \vspace{-0.2in}
\end{figure}

\subsubsection{Results of institution version}

We evaluate the performance of the institutional version of the FinGPT search agent against some commonly used LLMs. The results for EMIR, ESMA, Link Retrieval, and NER are reported in Table~\ref{tab:model_comparison_2}, and the results for FinanceBench are shown in Fig.~\ref{fig:5}. The observations are as follows.

The FinGPT Search Agent outperforms all the baselines across all datasets, achieving 0.85 and 0.98 on EMIR abbreviations and definitions respectively. Similar observations can be made on ESMA, Link Retrivel, NER, and FinanceBench. The improvement is more pronounced in Link Retrieval, nearly doubling the accuracy achieved by GPT-4o. One explanation might be that link retrieval often requires access to external web pages to identify links. The FinGPT Search Agent is capable of effectively retrieving online information and utilizes RAG to further improve accuracy. These results demonstrate the strong performance of the FinGPT Search Agent compared to the best current large language models.

\begin{table*}[t]
\centering
\caption{Mean $\pm$ standard deviation of the response time of the FinGPT search agent and the baselines in milliseconds (ms).}
\label{tab:response_times}
\begin{tabular}{lcc}
\toprule
\textbf{Model} & \textbf{Financial Indices} & \textbf{Web Questions} \\ 
\midrule
\textbf{GPT-3.5 Turbo} \cite{openai2023gpt3.5} & $\textbf{2925.9} \pm \textbf{1471.5}$ & - \\ 
\textbf{GPT-4} \cite{openai2023gpt4} & $12512.8 \pm 4077.3$ & - \\ 
\textbf{GPT-4o} \cite{openai2024gpt4o} & $8720.7 \pm 2209.9$ & - \\ 
\textbf{Mistral} \cite{mistral7b} & $7886.6 \pm 1483.2$ & - \\ 
\textbf{LLaMA 3.1} \cite{llama3modelcard} & $10223.1 \pm 2913.6$ & - \\ 
\textbf{LLaMA 3} \cite{llama3modelcard} & $10824.2 \pm 1634.2$ & - \\ 
\textbf{FinGPT Search Agent (GPT-3.5 Turbo as base)} & $\textbf{7658.0} \pm \textbf{1166.2}$ & $2244.9 \pm 526.4$ \\ 
\textbf{FinGPT Search Agent (GPT-4o as base)} & $8016.9 \pm 1318.8$ & $\textbf{2087.8} \pm \textbf{717.4}$ \\ 
\textbf{FinGPT Search Agent (LLaMA 3.1 as base)} & $7768.1 \pm 1007.5$ & $2509.7 \pm 993.4$ \\ 
\bottomrule
\end{tabular}
\vspace{-0.12in}
\end{table*}

The results in the FinanceBench dataset shown in Fig.~\ref{fig:5} shows LLaMA-3-RAG generally outperforming LLaMA-3-8B, particularly in domain-relevant and novel-generated questions. This demonstrates the effectiveness of RAG. Additionally, the Customized FinGPT Search Agent (with GPT-4o as the base) outperforms GPT-4o across all three types of questions. This justifies incorporating RAG into the Customized FinGPT Search Agent.

\subsection{Efficiency of FinGPT Search Agent (RQ2)}

We evaluate the efficiency of the individual version of FinGPT Search Agent on its response time. We compare the response times to the baseline models. We test six base models and use three generally most powerful ones as the base models for the search agent for this task. We measured the response times of the FinGPT search agent and the base models on 20 queries, with an average query length of 67.3 characters. The results are reported in Table~\ref{tab:response_times}. Base models cannot access the Internet, and thus were not tested for web questions. 

Despite the additional computational overhead, the FinGPT search agent maintains a reasonable response time, producing results in several seconds. The search agent's response time averages 7658.0 ms for financial indices and 2244.9 ms for web questions when using GPT-3.5 Turbo, representing the fastest response time. GPT-3.5 Turbo itself achieved 2925.9 ms on financial indices, outperforming all other models on this task. Notably, the search agent using GPT-4o and LLaMA 3.1 as base models achieved an average response time of 8016.9 ms and 7768.1 ms respectively for financial indices, outperforming the base models themselves. A possible reason is that RAG can provide more relevant context, guiding the model to more efficient response generation.

\section{Conclusion and Future Work}

In this paper, we presented two customized FinGPT search agents: one for individual users and another for institutional users. The individual version focuses on providing tailored financial planning and advice, integrating user preferences, local files, and real-time web data through RAG. The institutional version is designed to analyze proprietary datasets, leveraging continual pretraining, finetuning, and RAG to meet the specific needs of financial institutions. Both agents operate locally, ensuring data security. The experimental results demonstrate that both versions of the FinGPT search agents outperform existing LLMs in terms of accuracy and data freshness, indicating their potential for real-world applications.

Future work will explore further optimization of better information retrieval methods, improved compliance with regulatory standards, and enhanced user experiences to make it more user-centric.

\begin{acks}

Felix Tian, Ajay Byadgi, Daniel Kim, and Xiao-Yang Liu Yanglet acknowledge the support from NSF IUCRC CRAFT center research grant (CRAFT Grant 22017) for this research. The opinions expressed in this publication do not necessarily represent the views of NSF IUCRC CRAFT. Felix Tian and Xiao-Yang Liu Yanglet acknowledge the support from Columbia's SIRS and STAR Program, The Tang Family Fund for Research Innovations in FinTech, Engineering, and Business Operations.  

\end{acks}

%%
%% The next two lines define the bibliography style to be used, and
%% the bibliography file.
\bibliographystyle{ACM-Reference-Format}

\bibliography{references}

\end{document}